\documentclass{article}

\usepackage{PRIMEarxiv}

\usepackage[utf8]{inputenc} 
\usepackage[T1]{fontenc}    
\usepackage{hyperref}       
\usepackage{url}            
\usepackage{booktabs}       
\usepackage{amsfonts}       
\usepackage{nicefrac}       
\usepackage{microtype}      
\usepackage{lipsum}
\usepackage{fancyhdr}       
\usepackage{graphicx}       
\graphicspath{{media/}}     
\usepackage{amsmath}
\usepackage{siunitx}
\usepackage[table]{xcolor}
\newcommand{\best}[1]{\textbf{#1}}

\newcounter{hypothesis}

\refstepcounter{hypothesis}\label{hyp:less-Bias}
\refstepcounter{hypothesis}\label{hyp:societal-stereotypes}
\refstepcounter{hypothesis}\label{hyp:context-mitigates-bias}
\refstepcounter{hypothesis}\label{hyp:Western-content}
\newcommand{\tpms}[1]{\raisebox{0.0ex}{\scriptsize{\ensuremath{\,\pm\,}#1}}} 

\title{Revealing Potential Biases in LLM-Based Recommender Systems in the Cold Start Setting
}

\author{Alexandre Andre
    \thanks{Both authors contributed equally to this research.}\\
    University of Pennsylvania \\
    Philadelphia \\
    \texttt{aandre1@seas.upenn.edu}
    \And 
    Gauthier Roy \footnotemark[1] \\
    Georgia Institute of Technology \\
    Atlanta \\
    \texttt{gauthier.roy@etu.utc.fr} 
    \AND 
    Eva Dyer\\
    University of Pennsylvania\\
    Philadelphia\\
    \texttt{eva.dyer@seas.upenn.edu} \\
    \And 
    Kai Wang \\
    Georgia Institute of Technology \\
    Atlanta \\
    \texttt{kwang692@gatech.edu}
}

\begin{document}
\maketitle

\begin{abstract}
    Large Language Models (LLMs) are increasingly used for recommendation tasks due to their general-purpose capabilities. While LLMs perform well in rich-context settings, their behavior in cold-start scenarios, where only limited signals such as age, gender, or language are available, raises fairness concerns because they may rely on societal biases encoded during pretraining. We introduce a benchmark specifically designed to evaluate fairness in zero-context recommendation. Our modular pipeline supports configurable recommendation domains and sensitive attributes, enabling systematic and flexible audits of any open-source LLM. Through evaluations of state-of-the-art models (Gemma 3 and Llama 3.2), we uncover consistent biases across recommendation domains (music, movies, and colleges) including gendered and cultural stereotypes. We also reveal a non-linear relationship between model size and fairness, highlighting the need for nuanced analysis.
\end{abstract}


\section{Introduction}

Recommendation systems shape the digital experiences of billions of users, guiding what we read, watch, learn, and purchase. These systems play an essential role in helping users navigate large catalogs and discover new content or opportunities. From media and shopping to education and career planning, recommender systems are embedded in many high-stakes domains, making their performance, fairness, and trustworthiness critical.

The recent success of Large Language Models (LLMs) \cite{wei2022finetunedlanguagemodelszeroshot, brown2020languagemodelsfewshotlearners} has opened new avenues for building general-purpose recommendation systems. These models can generate recommendations directly from prompts, using their broad knowledge and linguistic flexibility to understand user goals and suggest relevant items \cite{delcluze2025text2playlistgeneratingpersonalizedplaylists}. This ability is especially valuable in scenarios where item descriptions, tags, or other textual information are available \cite{geng2023recommendationlanguageprocessingrlp}. However, with this opportunity comes a new set of challenges. Prior work has shown that LLM-based recommendation can amplify social biases, especially when user profiles include sensitive attributes such as gender, age, or language \cite{Zhang_2023}.

While existing work has begun to explore fairness in LLM-driven recommendation \cite{Zhang_2023, wan2024faireval, wang2024cfairllm}, a critical gap remains: the cold start setting. In this scenario, platforms have little to no interaction history for a user and must often rely on limited signals to generate recommendations. This creates a risky dynamic where LLMs may overfit to stereotypes or social priors encoded during pretraining. Left unchecked, these behaviors can lead to biased suggestions that shape user behavior in ways that reinforce inequality or discourage engagement.

In this work, we introduce a new benchmark and pipeline for assessing bias in cold start recommendation scenarios. Our framework is designed for the cold start setting, where only sensitive attributes are available. It features a modular pipeline with configurable datasets and sensitive attributes that allows practitioners to systematically evaluate any open-source LLM hosted on the Hugging Face Hub. In contrast to prior work \cite{Zhang_2023, wan2024faireval, wang2024cfairllm}, we introduce new recommendation domain with college recommendation, where bias may also be critical to characterize.

We demonstrate the utility of our benchmark with case studies using two state-of-the-art LLMs, Gemma 3 and Llama 3.2. Our experiments uncover biases in model outputs across domains, including music, movie, and college recommendation. For example, we provide evidence that across domains there are complex, non-linear relationships between model size and bias, highlighting the nuanced and domain-specific nature of fairness in LLM recommendations. Additionally, we show that LLMs tend to exhibit a bias toward Western content, often recommending predominantly Western-produced movies to neutral users, thereby aligning their suggestions with those made to Western-identified users. These results demonstrate how our pipeline supports flexible, reproducible audits of model behavior and facilitates comparative analysis across model architectures and recommendation domains.\\

Our key contributions are:
\begin{itemize}
    \item A benchmark for cold-start recommendation that isolates and evaluates model behavior when only sensitive user attributes are provided. The benchmark is implemented as a modular and extensible pipeline that supports a variety of open-source LLMs with configurable recommendation domains and sensitive attributes.
          
    \item Case studies providing empirical evidence that LLMs reproduce societal biases, including gender and cultural stereotypes, and revealing a complex, non-linear relationship between model scale and fairness.

\end{itemize}

\section{Related Works}

\paragraph{LLM-Based Recommender Systems in Cold-Start} Recent advances in LLMs have enhanced recommender systems, particularly in cold-start settings \cite{geng2023recommendationlanguageprocessingrlp}. Seminal work by \cite{10.48550} showed that LLMs can perform well without prior user-item interactions by generalizing from textual data. This has inspired the development of specialized architectures. For example, the TALLRec framework demonstrated effective and efficient tuning for recommendation tasks \cite{Bao_2023}, while the FilterLLM architecture \cite{liu2025filterllmtexttodistributionllmbillionscale} introduced a Text-to-Distribution approach, achieving significant efficiency gains in Alibaba’s cold-start recommendation system.

\paragraph{Bias and Fairness in LLMs} Various datasets assess bias in LLMs. For stereotypical biases, benchmarks like StereoSet \cite{nadeem2021stereoset} and the Bias Benchmark for QA (BBQ) \cite{parrish2022bbqhandbuiltbiasbenchmark} are used to probe harmful social stereotypes. Toxicity is evaluated using datasets such as RealToxicityPrompts \cite{gehman2020realtoxicityprompts} for explicit content and ToxiGen \cite{hartvigsen-etal-2022-toxigen} for more implicit forms of toxicity. Broader ethical alignment is measured by comprehensive benchmarks like ETHICS \cite{hendrycks2021ethics} and through datasets derived from real-world user interactions like Eagle \cite{kaneko2024eagleethicaldatasetgiven}. Methodologically, counterfactual analysis \cite{liang2021towards} helps identify biases in LLM outputs, with recent frameworks now being developed to formally certify this fairness \cite{chaudhary2025certifyingcounterfactualbiasllms}.

\paragraph{Bias and Fairness in Recommender Systems} Recommender systems can be biased in terms of popularity, exposure, and demographics \cite{abdollahpouri2017controlling, diaz2020evaluating, burke2017multisided}. Fairness metrics focus on disparities in recommendation quality and exposure, with measures like coverage, diversity, and popularity bias \cite{ge2010beyond, steck2018calibrated}. Metrics like disparate impact and demographic parity are also used to assess fairness \cite{burke2017multisided, yao2017beyond}. Recent surveys have sought to systematize the field by providing detailed taxonomies of fairness concepts and mitigation strategies \cite{jin2023surveyfairnessawarerecommendersystems}.

\paragraph{Bias and Fairness in LLM-Based Recommender Systems} FairEval \cite{wan2024faireval} integrates personality and demographic attributes to assess bias, using metrics like the Personality-Aware Fairness Score (PAFS). CFaiRLLM \cite{wang2024cfairllm} evaluates fairness using the same metrics as we use. These frameworks, similar to ours, adapt LLM bias evaluation datasets to the recommender context, probing for biases based on user identity through techniques like controlled prompt variations.

\section{Problem Statement}

In a setting where a user asks a chatbot for recommendations, we aim to automatically detect whether the LLM introduces bias based on the user's sensitive attributes, potentially discriminating  certain user categories.

\subsection{Challenges}

Recommendation systems that leverage LLMs face significant challenges due to the rapid evolution of models and the diverse recommendation domains of application. To remain effective and sustainable, such systems must be designed with flexibility and robustness at their core.

\paragraph{Flexibility} There is a need for a modular and flexible pipeline that can adapt to ongoing rapid changes in LLM. Indeed, the pipeline must ensure compatibility across different model versions without requiring substantial redesign. Additionally, it should support varying datasets and attribute configurations, enabling integration across multiple recommendation domains. This flexibility is especially important for LLM providers who may wish to add or remove specific sensitive attributes of interest.

\paragraph{Robustness} The pipeline must be robust and automated to reduce manual intervention. This includes the ability to handle prompts that are generalizable across different settings, minimizing the need for frequent tuning or rewriting. Furthermore, the pipeline should be capable of parsing and standardizing diverse LLM output formats, ensuring consistent performance regardless of the model’s response style.

\subsection{Task}
In the context of LLM-based recommendation systems in the cold start setting, a new user typically expect to be recommended a small number of highly relevant items, rather than retrieve or generate long, exhaustive lists. To address this expectation, we introduce a re-ranking task as the central focus of our framework.

We formalize re-ranking the following way. We provide a catalog of $ N $ items to the LLM and ask it to select and rank a small subset of $ k $ items (e.g., 5, 10, or 20) to match the user's taste. Given a large-scale dataset of size \begin{math} N_{\text{real}} \end{math}, restricting the LLM's access to a much smaller subset (\begin{math} N \ll N_{\text{real}} \end{math}) is motivated both by technical constraints, the LLM's limited context window, and by the task design, the production of a small, personalized, ordered list of recommendations. 

We argue that our evaluation is better suited for cold-start scenarios and more effective at capturing bias compared to the task used in benchmarks like \cite{Zhang_2023, wan2024faireval, wang2024cfairllm}. For example in FaiRLLM \cite{Zhang_2023}, the LLM is prompted to generate an ordered list of recommended items associated with a specific artist, movie director, or actor, providing strong contextual cues that undermine the cold-start setting.

\section{Methodology}

\begin{figure*}
    \centering
    \includegraphics[width=\linewidth]{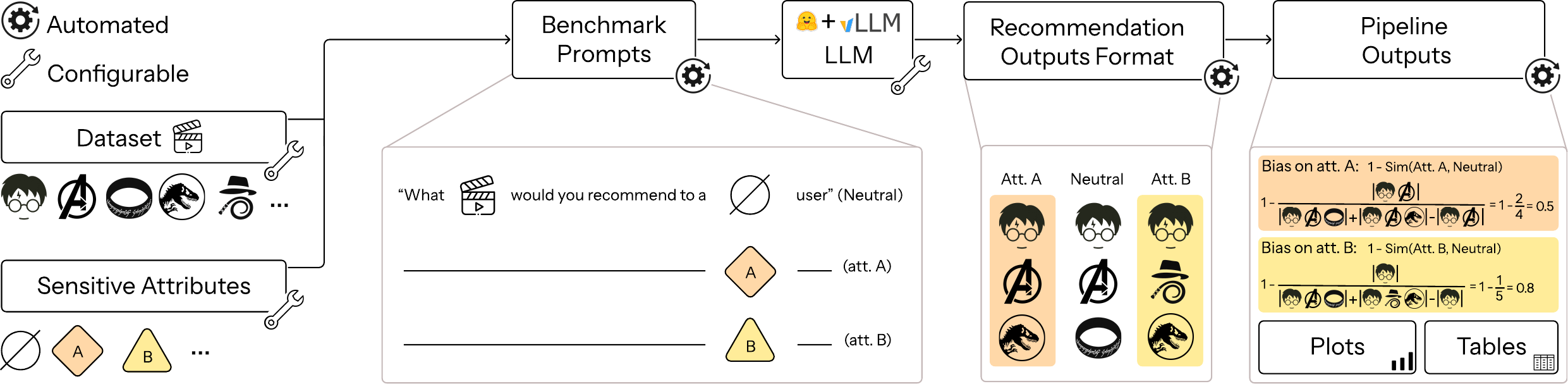}
    \caption{\textbf{Overview of the benchmark pipeline for measuring bias across datasets and sensitive attributes.} The user provides datasets, sensitive attributes, and the LLM to be tested. From these inputs, the pipeline automatically generates benchmark prompts that vary by sensitive attribute (e.g., attribute A, attribute B, and neutral). The LLM is queried with these prompts and its outputs are formatted into recommendation sets per attribute condition. These recommendation sets from different sensitive attributes are compared against the neutral baseline using a similarity metric (here, IOU) to compute bias scores. Finally, the results are compiled and can be directly inspected as raw values or visualized through automatically generated plots and tables.}
    \label{fig:pipeline}
\end{figure*}

Our primary objective is to provide an automated and modular pipeline, evaluating the bias of LLMs' ranked suggestions on recommendation datasets. In this section, we first present the benchmark pipeline, followed by the formal details of the metrics used, and then the datasets.

\subsection{Pipeline}

The pipeline, depicted in Figure~\ref{fig:pipeline}, operates in a sequence of automated steps. It starts by taking a \textit{dataset} and a list of \textit{sensitive attributes} to generate \textit{benchmark prompts} tailored to neutral users and users with specific attributes. A key feature is our use of vLLM \cite{kwon2023efficient}, which enables efficient, scalable inference on any open-source LLM from the Hugging Face Hub, a significant extension over prior work focused on closed APIs \cite{Zhang_2023, wang2024cfairllm}.

The model's responses are then systematically organized into a structured \textit{recommendation outputs format}. Finally, the pipeline computes its \textit{outputs}, including bias scores derived from pairwise comparisons between the recommendations for neutral and attribute-specific users. It generates from the scores plots and tables that show the bias in relation to sensitive attribute. This end-to-end automation, combined with a configurable dataset, sensitive attribute list, and LLM integration, makes our pipeline a powerful and extensible tool for conducting robust fairness analysis.

\subsection{Metrics}

Formally, let us consider an ordered list \begin{math} \mathcal I_\text{Neu.}^k \end{math} of $ k $ items recommended by the LLM for a neutral user (i.e., a user without any attributes). In a counterfactual setting, \begin{math} \mathcal I_\text{Neu.}^k \end{math} is compared to \begin{math} \mathcal I_a^k \end{math}, the recommendations generated for a user with the sensitive attribute $ a $.

We can define \begin{math} \text{Sim}(\mathcal I_a^k, \mathcal I_\text{Neu.}^k) \end{math}, which evaluates how close the two lists are depending on the  similarity measure chosen. This assesses the impact of the sensitive attribute $ a $ on the recommendations produced by the LLM. The score will range between 0 and 1 and the two lists are identical when the score is 1.

In an ideal scenario, a perfectly unbiased LLM would yield \begin{math} \allowbreak \text{Sim}(\mathcal I_a^k, \mathcal I_\text{Neu.}^k) = 1 \end{math} for all sensitive attributes. However, such behavior may also suggest that the model does not personalize its recommendations, an issue which could negatively affect performance. These trade-offs raise concerns, which we address in the discussion part of the paper.

We define the bias in the LLM as the complement of the similarity score. Note that we also use the term divergence, associated with the similarity measures, to express the bias with respect to each measure. Formally, we define the bias with respect to a specific attribute $ a $ as \begin{math} B_{\text{Sim}}^k(a) = 1 - \text{Sim}(I_a^k, \mathcal I_\text{Neu.}^k) \end{math}. Then, a value of $ B^k $ close to zero indicates a small bias with respect to this attribute. The bias measure can be used for comparison across different attributes; for instance, \begin{math} B_k(a) < B_k(b) \end{math} would indicate that the LLM is less biased with respect to attribute $ a $ than attribute $ b $.

We retain the similarity metrics used in \cite{Zhang_2023}.
The Jaccard similarity, subsequently referred as IOU, treats the lists as unordered. SERP takes into account the order of the items in the list, giving more weight to items that appear in both lists and are ranked higher. PRAG considers the ranked lists and the relative order of items in both lists.

Depending on the similarity measure used, we formally define \(B^k(a)\):
$$
B_{\text{IOU}}^k(a) = 1 - \frac{\left| \mathcal I_a^k \cap \mathcal I_{\text{Neu.}}^k\right|}{\left| \mathcal I_a^k \right| + \left| \mathcal I_{\text{Neu.}}^k\right| - \left| \mathcal I_a^k \cap \mathcal I_{\text{Neu.}}^k\right|}
$$

\medskip

$$
B_{\text{SERP}}^k(a) = 1-\sum_{i \in  \mathcal I_a^k} \frac{2 \cdot \mathbf{1}_{i_1 \in \mathcal I_\text{Neu.}^k } \cdot \left(k - r_a(i) +1\right)}{k(k+1)}
$$
\medskip
$$     
B_{\text{PRAG}}^k(a) = 1 - \sum_{\substack{i_1, i_2 \in I_a^k \\ i_1 \neq i_2}}\frac{\mathbf{1}_{i_1 \in \mathcal I_\text{Neu.}^k} \cdot \mathbf{1}_{r_\text{Neu.}(i_1)< r_\text{Neu.}(i_2)} \cdot \mathbf{1}_{r_a(i_1)< r_a(i_2)} }{k(k+1)} 
$$

\noindent With \(\mathbf{1}\) indicator function, \(r_a(i)\) (respectively \(r_\text{Neu.}(i)\)) rank of the item \(i\) in \(I_a^k\) (respectively \(I_\text{Neu.}^k\)).\\

A useful perspective is to interpret $ I_a^k $ as the top-$ k $ items sampled from the distribution \begin{math} P(\text{item}\mid a) \end{math}. Here, $ P $ represents, for a specific LLM, the preference distribution over items in the recommendation catalog, conditioned on the user having attribute $ a $. Similarly, $ I_{\text{Neu.}}^k $ can be seen as the top-$ k $ items drawn from $ P(\text{item}) $, the distribution without conditioning on any attribute. Thus, $ B^k(a) $ measures how the distribution $ P(\text{item} \mid a) $ differs from $ P(\text{item}) $ for the top-$k$ items, reflecting how a specific sensitive attribute shapes the likelihood of the most probable items being recommended.

Note that the probability distribution $ P $, in our case, is restricted, as we provide the LLM with a list of $ N $ items to select from and order to match the user's preferences. This list is what we refer to as the dataset.

\subsection{Dataset}
\label{sec:dataset}

For the datasets used in our benchmark, we retain the music and movie recommendation domains covered by \cite{Zhang_2023, wang2024cfairllm, wan2024faireval} and introduce a new recommendation domain: colleges. We believe that college recommendations, compared to cultural product recommendations (e.g., music and movies), raise additional ethical concerns, as they have the potential to influence a person's educational and career opportunities. This makes ensuring fairness in LLM-generated recommendations within this domain even more critical.

Since we are working in a re-ranking setup, we limit the number of items provided to the LLM. Typically, recommendation systems re-rank lists of 1,000 items, but preliminary experiments revealed that the LLM’s context window could not effectively keep in memory all 1,000 items. As a result, we decided to reduce the number of items to 500 to maintain variability in the item list while making the task more manageable for the LLM. The music and movie datasets are obtained via APIs, so their replicability is not guaranteed. However, we provide the lists of 500 items for all datasets to ensure reproducible results and allow the community to test their LLMs on these datasets without needing to collect data. The framework also offers users the flexibility to extend the benchmark and add additional item lists for recommendation in domains of their interest.

We provide technical details on how the three datasets used in the benchmark were created.
\begin{itemize}
    \item Movie: Using the \href{https://developer.themoviedb.org/}{Movie Database (TMDB) API}, highest-rated movies from IMDb and retain their titles in the English-language version.
    \item Music: Using the \href{https://developer.spotify.com}{Spotify Web API} and \href{https://spotipy.readthedocs.io/en/2.25.1/}{Spotipy}, we extracted a list of popular songs ranging from the 1970s to the 2010s. We took the first 100 songs from the playlist Acclaimed Music of every decades (e.g., "Top Songs of the 2010s – Acclaimed Music"), which features critically acclaimed tracks based on aggregated rankings from music critics' lists compiled by the website \href{http://www.acclaimedmusic.net}{Acclaimed Music}. We reformatted the selected songs to follow this structure: \textit{[Song title] by [Artist name]}.
    \item College: Using university names from \href{https://www.kaggle.com/datasets/tahirrfarooqq/2023-qs-world-university-ranking}{2023 QS World University Rankings dataset}, available on Kaggle, which initially rank more than 1,400 institutions.
\end{itemize}

\section{Benchmark Generation and Findings from the Pipeline Outputs}

We begin our investigation by leveraging the counterfactual framework presented in previous section, which involves comparing recommendations generated from prompts that either include or omit specific sensitive user attributes (e.g., gender, nationality). This approach allows us to systematically probe how LLMs respond when only sensitive attributes are available, simulating a challenging cold-start recommendation scenario.

\subsection{Hypotheses to Investigate}
\label{sec:hypotheses}
To structure our investigation, we formulate the following \arabic{hypothesis} hypotheses.

\begin{itemize}
    \item \textit{\ref{hyp:less-Bias}: Larger LLMs exhibit less Bias.} This hypothesis confronts the conventional wisdom that "bigger is better" in AI \cite{kaplan2020scalinglawsneurallanguage}. We test whether increased model scale is a straightforward solution for fairness, or if it reveals a more complex relationship with potential trade-offs between a model's capabilities and its biases.

    \item \textit{\ref{hyp:societal-stereotypes}: LLMs replicate societal stereotypes.} This hypothesis suggests that LLMs, acting as mirrors of their training data, are likely to reproduce well documented societal stereotypes. We specifically test for gender based stereotypes in movie recommendations, a domain where such biases are known to be prevalent \cite{10.1145/3696410.3714528, sakib2024challenging, https://doi.org/10.5281/zenodo.10469839}.

    \item \textit{\ref{hyp:context-mitigates-bias}: Adding context to a user mitigates bias.} This hypothesis explores a potential mitigation for the tendency of LLMs to default to stereotypes in low information, cold start scenarios. We test the idea that providing task relevant user preferences can override the model's reliance on sensitive attributes, thereby reducing bias.

    \item \textit{\ref{hyp:Western-content}: LLMs are biased towards Western content.} This hypothesis scrutinizes the default cultural lens of LLMs. We move beyond simple nationality prompts to reveal the model's assumed '\textit{neutral}' user, exposing the depth of its bias towards Western content when no specific culture is mentioned.
\end{itemize}

These hypotheses guide the experimental analysis presented in the following subsection, where we use our benchmark framework to gather evidence supporting or refuting each claim.

\subsection{Experiments}

\subsubsection{Setup}

The experiments presented in this section leverage the benchmark pipeline detailed previously. Four instruction-tuned LLMs were selected for evaluation: 
\begin{itemize}
    \item \texttt{meta-llama/Llama-3.2-3B-Instruct} (Llama 3.2 3B).
    \item \texttt{google/gemma-3-1b-it} (Gemma 3 1B).
    \item \texttt{google/gemma-3-4b-it} (Gemma 3 4B).
    \item \texttt{google/gemma-3-12b-it} (Gemma 3 12B). 
\end{itemize}

For each experiment run, models were presented with a catalog comprising 500 items sourced from either the College, Music, or Movie datasets. They were then prompted to select and rank the top 20 items ($ k=20 $) tailored to a user with a sensitive attribute or a neutral (i.e., without attribute). To ensure the robustness and reliability of our quantitative findings, all reported metrics and percentages represent the average results obtained across 5 independent generation seeds. Error bars depicted in the plots correspond to the standard deviation calculated over these 5 seeds, providing insight into the variability of the model responses.

\subsubsection{Initial Model Comparison and Selection}

A preliminary comparison between Gemma 3 4B and Llama 3.2 3B was conducted to inform model selection for further hypothesis testing. This comparison utilized overall mean IOU, SERP, and PRAG Divergence metrics, aggregated across all sensitive attributes within each dataset (Table~\ref{tab:iou_divergence_gemma_vs_llama}).

On the College dataset, Gemma showed higher mean divergence for each metric, indicating greater output variation from the baseline. On the Music dataset, Llama had significantly higher mean IOU and PRAG divergence (0.46 IOU, 0.37 PRAG) than Gemma (0.27 IOU, 0.20 PRAG). For the Movie dataset, both models had similar IOU and SERP divergence, but Gemma had lower PRAG divergence (0.46 vs. 0.50 for Llama).

To finalize the decision, a qualitative analysis of raw model outputs revealed substantial instability in Llama’s responses, particularly repetition artifacts (e.g., repeating the same song multiple times). This instability likely inflated the divergence score, as erratic outputs deviated significantly from the neutral baseline and coherent attribute-influenced lists. In contrast, Gemma generated more stable and coherent outputs. Given the need for reliable, interpretable results for bias analysis, Gemma 3 was selected as the primary focus for subsequent hypothesis testing.

\begin{table}
\centering
\renewcommand{\arraystretch}{1.2}
\caption{\textbf{Mean $\pm$ standard deviation of metric divergences for Gemma 3 (4B) and Llama 3.2 (3B) across the College, Music, and Movie datasets.} Best results per dataset are highlighted.}
\label{tab:iou_divergence_gemma_vs_llama}

\begin{tabular}{l c c c c c c}
    \toprule
    & \multicolumn{2}{c}{\textbf{College}} 
    & \multicolumn{2}{c}{\textbf{Music}} 
    & \multicolumn{2}{c}{\textbf{Movie}} \\
    \cmidrule(lr){2-3} \cmidrule(lr){4-5} \cmidrule(lr){6-7}
    \textbf{Metric} & Gemma & Llama & Gemma & Llama & Gemma & Llama \\
    IOU   & .55 \tpms{.12} & \best{.50} \tpms{.14} & \best{.27} \tpms{.09} & .46 \tpms{.15} & .55 \tpms{.09} & \best{.54} \tpms{.11} \\
    SERP  & .83 \tpms{.03} & \best{.80} \tpms{.05} & .77 \tpms{.02} & \best{.76} \tpms{.05} & .81 \tpms{.03} & \best{.80} \tpms{.04} \\
    PRAG  & .47 \tpms{.09} & \best{.46} \tpms{.12} & \best{.20} \tpms{.06} & .37 \tpms{.14} & \best{.46} \tpms{.08} & .50 \tpms{.13} \\

    \bottomrule
\end{tabular}
\end{table}

\subsubsection{Hypothesis Testing with Gemma Models}
We evaluate the hypotheses using the Gemma models.

\paragraph{\ref{hyp:less-Bias}: Larger LLMs exhibit less bias}
This hypothesis suggests that larger LLMs, with their increased capacity, might better understand and mitigate biases. However, our comparison of overall mean IOU, SERP, and PRAG Divergence across Gemma 1B, 4B, and 12B models (Table~\ref{tab:h2_gemma_size_bias}) reveals a more complex, non-monotonic relationship.

The Gemma 3 4B model consistently achieved the lowest mean divergence on the Music and Movie datasets, indicating the least bias. In contrast, the largest model, Gemma 3 12B, showed higher divergence than the 4B model, suggesting that while it follows instructions well, it might over-emphasize sensitive attributes. For instance, it may prioritize '\textit{French}' items more aggressively than the 4B model, leading to higher divergence. The smallest model, Gemma 3 1B, even though showcasing low divergence on College, exhibited the highest IOU and PRAG Divergence on other datasets. We attribute this high divergence to lower task fidelity, as it struggled with complex re-ranking instructions or staying within the provided item list.

Thus, the 4B model appears to strike a balance: it reliably executes the task and is less sensitive to attributes compared to the 12B model. Increasing model size does not necessarily reduce bias and may instead trade off task reliability for greater sensitivity to input attributes.

\paragraph{\ref{hyp:societal-stereotypes}: LLMs replicate societal stereotypes}

\begin{figure}[htbp]
    \centering
    \includegraphics[width=0.6 \linewidth]{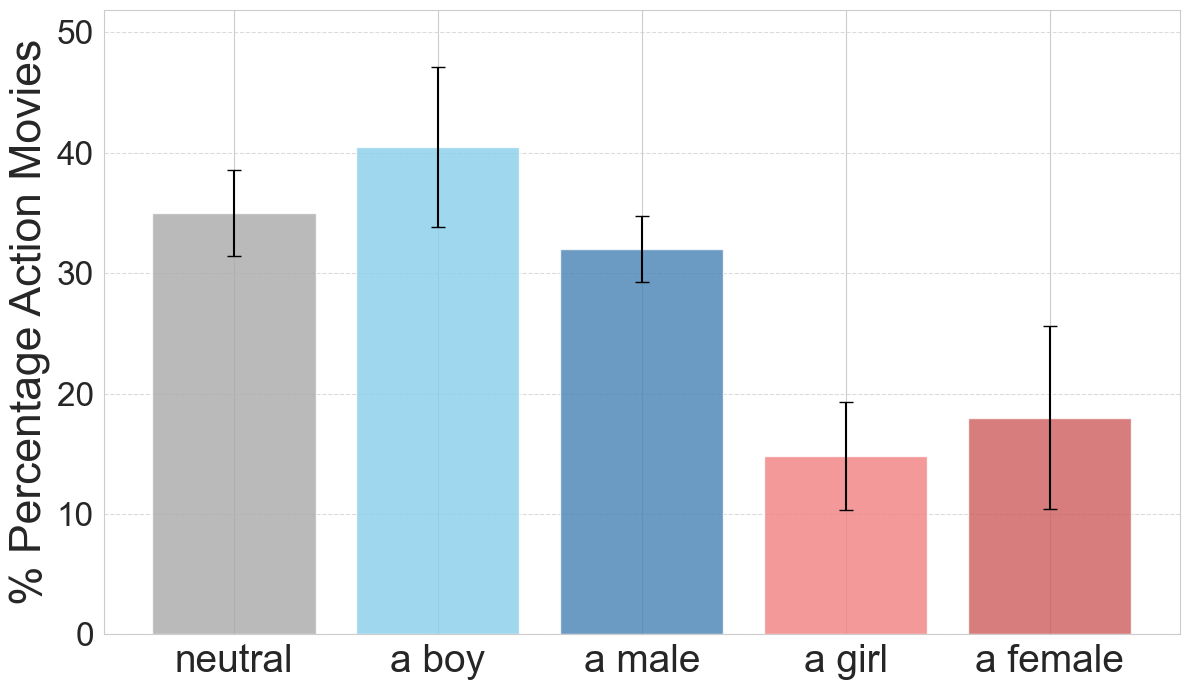} 
    \caption{\textbf{Ratio of Action Movies Recommended by Gemma 3 4B Across Gender Attribute.}}
\label{fig:h3_action_bias_4b}
\end{figure}

To test this hypothesis, we focused on potential gender stereotyping within the Movie dataset. Specifically, we analyzed the proportion of action movies appearing in the top-20 recommendations generated by Gemma 3 4B when prompted with different gender-related attributes (\textit{'a boy'}, \textit{'a girl'}, \textit{'a male'}, \textit{'a female'}) or with a neutral prompt. (Figure~\ref{fig:h3_action_bias_4b}).

The neutral user received approximately 35.0\% action movies. For the user with '\textit{a boy}' attribute, the proportion rose to 40.5\%, indicating a slight preference towards action films. Conversely, specifying '\textit{a girl}' dramatically reduced the action movie percentage to 14.8\%, and '\textit{a female}' similarly saw a reduction to 18.0\%. The '\textit{a male}' user yielded 32.0\%, slightly below the neutral baseline but significantly higher than '\textit{a girl}' or '\textit{a female}'.
This observed pattern of recommending more action movies to boys and significantly fewer to girls/females aligns with common societal stereotypes. \cite{https://doi.org/10.5281/zenodo.10469839} confirms these type of stereotypes in movie recommendations, showing that genres like sci-fi and thriller are more frequently recommended to \textit{male} users. 

These findings provide strong evidence supporting the hypothesis. 
This aligns with the results from H2, where the 12B model also showed higher overall divergence. The discrepancy between results for '\textit{boy}' or '\textit{girl}' versus '\textit{male}' or '\textit{female}' suggests the models are also sensitive to the specific phrasing used to denote gender, potentially reflecting different connotations or data associations learned for these terms.

\begin{table}[t]
\centering
\renewcommand{\arraystretch}{1.}
\caption{\textbf{Mean $\pm$ standard deviation of metric divergences for Gemma models of varying sizes (1B, 4B, 12B) across the College, Music, and Movie datasets.} Best results per dataset are highlighted.}
\label{tab:h2_gemma_size_bias}
\setlength{\tabcolsep}{5pt}
\begin{tabular}{l c c c c c c c c c}
    \toprule
    & \multicolumn{3}{c}{\textbf{College}} & \multicolumn{3}{c}{\textbf{Music}} & \multicolumn{3}{c}{\textbf{Movie}} \\
    \cmidrule(lr){2-4} \cmidrule(lr){5-7} \cmidrule(lr){8-10}
    \textbf{Metric} & 1B & 4B & 12B & 1B & 4B & 12B & 1B & 4B & 12B \\
    \midrule
    IOU   & .59 \tpms{.12} & \textbf{.55} \tpms{.12} & .65 \tpms{.07} & .73 \tpms{.20} & \textbf{.27} \tpms{.09} & .66 \tpms{.08} & .78 \tpms{.08} & \textbf{.55} \tpms{.09} & .75 \tpms{.07} \\
    SERP  & \textbf{.83} \tpms{.03} & .84 \tpms{.03} & .86 \tpms{.03} & .84 \tpms{.14} & \textbf{.77} \tpms{.02} & .85 \tpms{.03} & .86 \tpms{.04} & \textbf{.81} \tpms{.03} & .89 \tpms{.03} \\
    PRAG  & \textbf{.46} \tpms{.10} & .48 \tpms{.09} & .55 \tpms{.09} & .69 \tpms{.23} & \textbf{.20} \tpms{.06} & .48 \tpms{.09} & .72 \tpms{.09} & \textbf{.46} \tpms{.08} & .67 \tpms{.07} \\
    \bottomrule
\end{tabular}
\end{table}

\begin{figure}[htbp]
    \centering
    \includegraphics[width=0.6\linewidth]{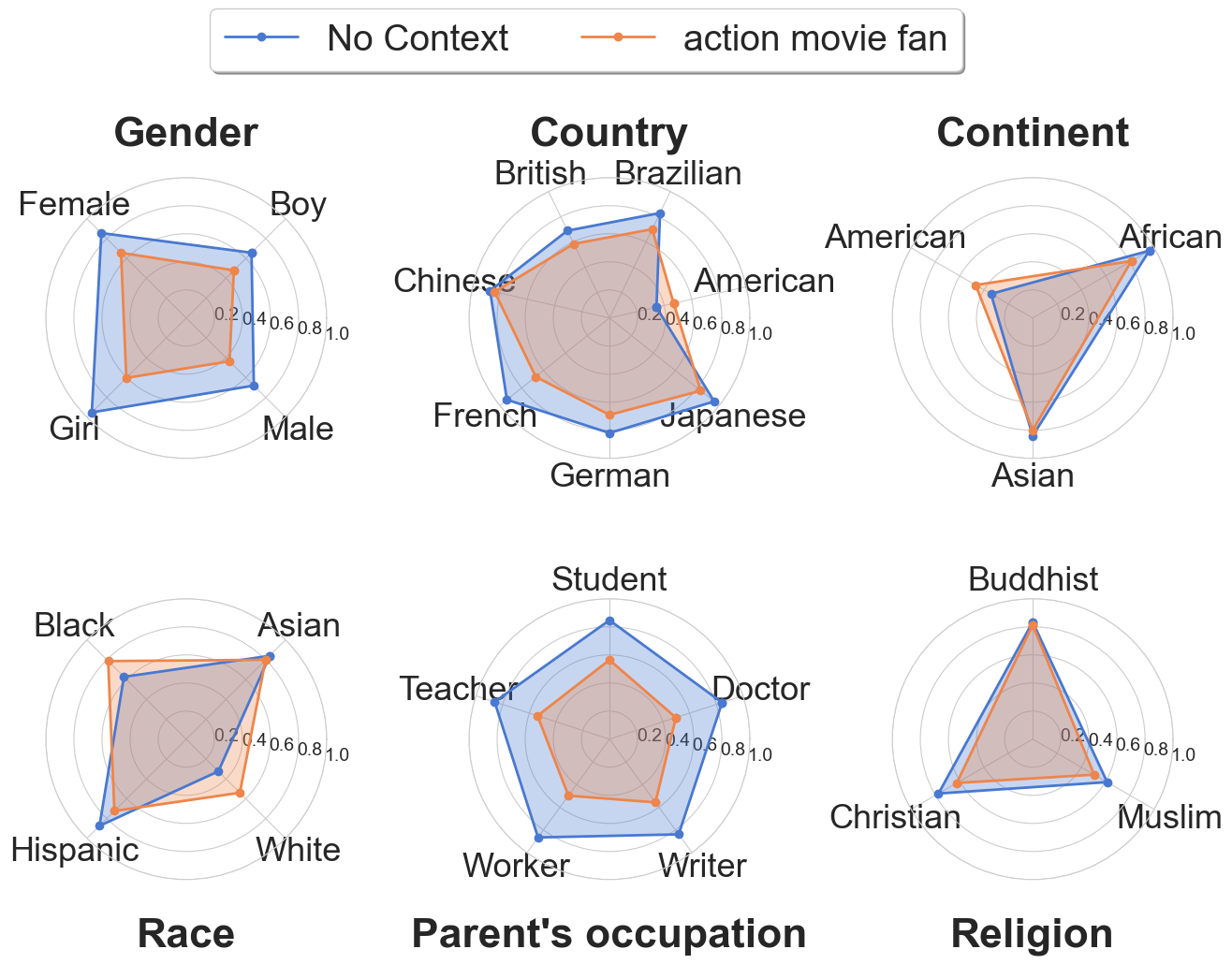} 
    \caption{\textbf{Impact of '\textit{Action Movie Fan}' Context on IOU Divergence in Gemma 3 12B (Movie Dataset)}}
    \label{fig:h4_spider_plot}
\end{figure}

\paragraph{\ref{hyp:context-mitigates-bias}: Adding context to a user  mitigates bias} 
This hypothesis proposes that incorporating relevant user preferences into the prompt, alongside sensitive attributes, can diminish the influence of those sensitive attributes, leading to reduced bias. We conducted this test using Gemma 3 12B on the Movie dataset, given its previously observed high sensitivity to attributes. Using IOU divergence for this analysis, we summarized the impact when the '\textit{action movie fan}' context is included, presenting the results in a spider plot (Figure~\ref{fig:h4_spider_plot}).

A general trend emerges: IOU divergence scores are lower across many sensitive attributes when the '\textit{action movie fan}' context is included. In Figure~\ref{fig:h4_spider_plot}, the orange line (with context) consistently lies closer from the center than the blue line (no context), reflecting the bias reduction. For example, divergence for gender attributes or parent's occupation is visibly reduced, with the effect especially pronounced for attributes with initially high divergence, such as '\textit{girl}'. This suggests that when the model receives a strong, task-relevant signal (e.g., preference for action movies), it prioritizes this over weaker, stereotype-driven signals, resulting in more similar recommendation lists across sensitive groups. Consequently, the visual evidence in Figure~\ref{fig:h4_spider_plot} provides strong support for the hypothesis: explicit, relevant user context can mitigate the bias observed in zero-context scenarios. 

\begin{figure}[htbp]
    \centering
    \includegraphics[width=0.75\linewidth]{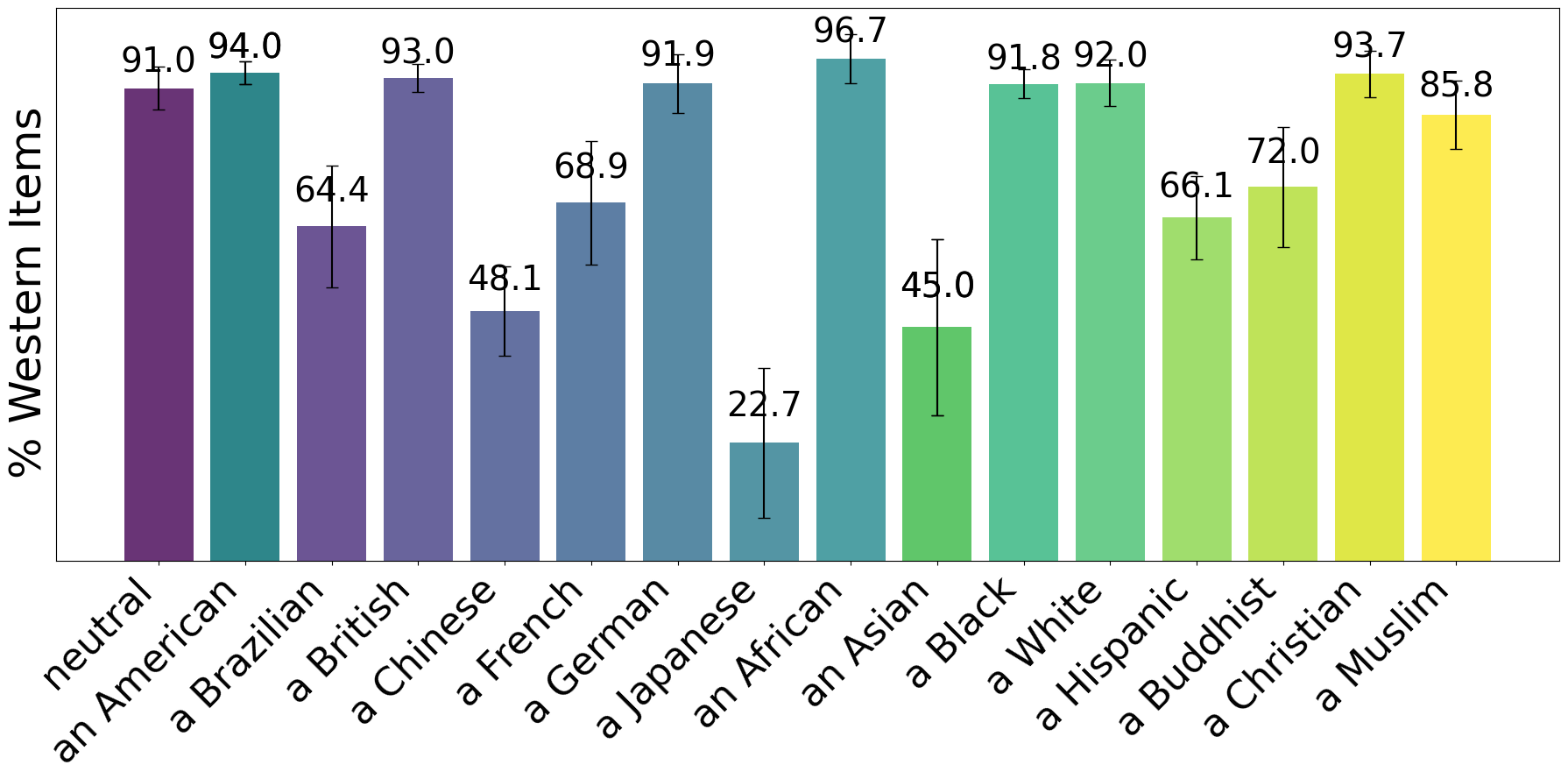} 
    \caption{\textbf{Ratio of Western Movies Recommended by Gemma 3 12B Across Different Attribute.}}
    \label{fig:h5_western_movie_bias}
\end{figure}

\paragraph{\ref{hyp:Western-content}: LLMs are biased towards Western content}
This hypothesis examines if LLM exhibit a broader bias favoring content from Western cultures. To evaluate the bias, we defined '\textit{Western}' content as primarily originating from North America, Europe, Australia, and New Zealand, and measured the percentage of such movies recommended by Gemma 3 12B across various personas (Figure~\ref{fig:h5_western_movie_bias}).

The results reveal a striking default preference: the neutral user, with no specified attributes, received recommendations that were mainly Western (91.3\%). This strongly suggests that, in the absence of specific user cues, the model defaults to recommending content aligned with Western cultures. Users with attribute of Western countries naturally maintained this high proportion (e.g., '\textit{an American}' ~94.0\%, '\textit{a German}' ~91.3\%, '\textit{a British}' ~89.3\%). While the model demonstrated some ability to adapt recommendations for non-Western users by including more non-Western films (e.g., '\textit{a Chinese}' received ~48.0\% Western content, '\textit{a Japanese}' received only ~22.0\%, '\textit{an Asian}' received ~45.3\%), the recommendations still leaned heavily towards Western films compakaired to an unbiased distribution. Even religious attributes, often associated with diverse global populations, resulted in high Western content percentages ('\textit{a Buddhist}' ~93.3\%, '\textit{a Muslim}' ~85.3\%). The extremely high percentage for '\textit{an African}' (96.0\%) appears anomalous and might reflect specific training data artifacts or misinterpretations.

These results strongly support the hypothesis, indicating a bias towards recommending Western content, in particular for the neutral case. This outcome is likely a direct reflection of the model's training data. Because the model was trained primarily on an English-language internet corpus, its knowledge base is inevitably skewed towards the cultural products of English-speaking, primarily Western, nations where much of this data originates. It would be interesting for future research to investigate these biases by prompting in different languages, or to explore whether similar levels of Western content bias are observed in models developed outside of Western contexts (e.g., DeepSeek \cite{deepseekai2025deepseekr1incentivizingreasoningcapability}).

\bigskip

\section{Discussion}

\paragraph{Limitations and Extensions} 
An alternative analysis could reverse the current setup, ranking users for a given item instead of ranking items for a user. This perspective could reveal model prioritization and biases more sharply, especially in fairness-sensitive applications like job recommendations. Comparing biases between item-to-user and user-to-item tasks would also illuminate how problem framing affects bias manifestation.

Looking ahead, a natural extension would be to tackle the retrieval task.  This could involve integrating online data by equipping the LLM with search capabilities, enabling it to recommend up-to-date items (e.g., newly released movies). This would provide a more flexible, dynamic dataset context.

Our experiments revealed a non-monotonic relationship between model size and bias, with Gemma 3 4B showing less divergence than both the 1B and 12B variants (\ref{hyp:less-Bias}). This motivates further investigation into even larger models (e.g., >70B parameters). It remains unclear whether bias would continue increasing, possibly due to heightened sensitivity to input, or whether larger models would develop a deeper, internalized understanding of fairness that mitigates bias.

This interplay between model capacity and bias invites exploration of explicit mitigation strategies. Prompt engineering is a promising direction: for example, explicitly instructing the model to '\textit{provide recommendations while avoiding unfair bias based on [sensitive attribute]}'. Evaluating such instructions would require assessing whether bias reduction sacrifices helpful personalization, or whether overcorrection occurs. Moreover, the ability to interpret and act on fairness instructions likely depends on model scale, with larger models potentially better suited to sophisticated bias mitigation via prompting or fine-tuning.

Finally, diversity metrics such as cross-entropy could also be explored to complement bias evaluation.

\paragraph{Ethical Consideration}
We acknowledge that bias assessment is not only a technical challenge but also an ethical imperative, necessitating a nuanced understanding of its societal implications. As we have discussed, bias in LLM-based recommendation systems might be reduced over time with more contextual information about the user. However, simply observing bias can also be used by platforms as a justification to further align models beyond default behavior \cite{das2024unveilingmitigatingbiaslarge}. A key question is whether certain biases are fair -- \textit{if the system recommends Italian movies to an Italian-speaking user, is that personalization or bias?} Ethical concerns arise when sensitive attributes, like gender, are used, as they risk reinforcing societal stereotypes. For example, \textit{if women are more likely to watch romantic movies, should the system continue to recommend them more of the same, potentially reinforcing gender-based preferences in a vicious circle?}

\section*{Acknowledgments}
We thank the Partnership for an Advanced Computing Environment (PACE) at the Georgia Institute of Technology for providing access to the server and computing resources that made this research possible. This paper is supported by NSF IIS-2403240, Schmidt Sciences AI2050 Fellowship, and CIFAR's Learning in Machines and Brains Program.

\bibliographystyle{unsrt}  
\bibliography{templateArxiv}

\end{document}